\documentclass[twocolumn,pra,aps,10pt,nofootinbib]{revtex4-2}   	% use "amsart" instead of "article" for AMSLaTeX format
\usepackage{geometry}                		% See geometry.pdf to learn the layout options. There are lots.
\usepackage{graphicx}				% Use pdf, png, jpg, or eps§ with pdflatex; use eps in DVI mode
\usepackage{tikz}
\usetikzlibrary{quantikz2}		
\usepackage{amssymb}
\usepackage{amsmath}
\usepackage{fancyhdr}
\usepackage{amsthm}
\usepackage{adjustbox}
\usepackage{hyperref}
\usepackage[capitalise]{cleveref}
\newtheorem{theorem}{Theorem}

\newtheorem{lemma}[theorem]{Lemma}

\theoremstyle{definition}

\DeclareFontFamily{U}{bbold}{}
\DeclareFontShape{U}{bbold}{m}{n}
 {
  <-5.5> s*[1.069] bbold5
  <5.5-6.5> s*[1.069] bbold6
  <6.5-7.5> s*[1.069] bbold7
  <7.5-8.5> s*[1.069] bbold8
  <8.5-9.5> s*[1.069] bbold9
  <9.5-11> s*[1.069] bbold10
  <11-15> s*[1.069] bbold12
  <15-> s*[1.069] bbold17
 }{}

\DeclareRobustCommand{\identity}{%
  \text{\usefont{U}{bbold}{m}{n}1}%
}

\begin{document}

\title{Implementing Clifford Gates on Stabilizer Codes via Measurement}
\author{Darren Banfield}
\affiliation{Department of Mathematics, Royal Holloway University of London, Egham Hill, Egham TW20 0EX, UK.}
\author{Heather Leitch}
\affiliation{Department of Mathematics, Royal Holloway University of London, Egham Hill, Egham TW20 0EX, UK.}
\author{Alastair Kay}
\affiliation{Department of Mathematics, Royal Holloway University of London, Egham Hill, Egham TW20 0EX, UK.}
\date{\today}

\bibliographystyle{ieeetr}

\begin{abstract}

Inspired by the code rewiring strategy of \cite{colladay_rewiring_2018}, we describe a method to use measurements and correction operations in order to implement the Clifford group in the code space of any stabilizer code, and we specify a sufficient set of conditions under which the distance of the code is preserved throughout. In particular this provides a method to implement a logical Hadamard-type gate within the 15-qubit Reed-Muller quantum code by measuring and correcting only two observables, providing the only non-transversal gate required for universality. Furthermore this approach is applicable to the toric code and quantum LDPC codes.

\end{abstract}

\maketitle

\section{Introduction}

Quantum computation offers the prospect of computation with much greater efficiency than classical computers for certain problems, as exemplified in \cite{shor_polynomial-time_1997,grover_fast_1996}. In practice, a major limitation of current quantum computing hardware is its susceptibility to error. Left unchecked, the noise resulting from these errors will damage the output of the computation. Much current research \cite{breuckmann_constructions_2016,hastings_fiber_2021,panteleev_degenerate_2021,breuckmann_balanced_2021,panteleev_asymptotically_2022,bravyi_high-threshold_2023} is aimed at correcting the errors, enabling the large-scale, reliable computations necessary for quantum computers to realise their potential.

Quantum error-correcting codes encode the data from a single (``logical") qubit in a block consisting of multiple physical qubits \cite{calderbank_good_1996,steane_multiple-particle_1997}. Operations are performed on the physical qubits, inducing the desired gate on the logical qubits. The fault-tolerant threshold theorem promises that there are families of codes that can achieve any desired accuracy provided the physical error rate is below an error threshold \cite{aliferis_quantum_2005}. Given the prevalence of errors, it is essential that this threshold be as high as possible, which can be achieved by designing new codes and optimising the implementation of their gates. In this paper, we present a new tool which allows any Clifford gate to be implemented on any additive quantum code.

Transversal gates, in which each physical qubit of a code is acted upon independently, are ideal for high threshold, low overhead, fault tolerant computation. Unfortunately, Eastin and Knill \cite{eastin_restrictions_2009} showed that it is not possible to construct a universal gate set for a quantum error-correcting code using only transversal gates. A common way to circumvent this issue is to implement almost all gates transversally, and then supplement them with magic state distillation \cite{knill_fault-tolerant_2004,bravyi_universal_2005,bravyi_magic_2012}. However, it is important to have a range of options available. One avenue that circumvents the high overheads induced by the ancilla preparation uses a pair of codes \cite{bombin_clifford_2011,colladay_rewiring_2018,anderson_fault-tolerant_2014,stephens_asymmetric_2008,bombin_quantum_2009,vuillot_code_2019} which between them implement a universal set of transversal gates. Switching between these two codes allows transversal gates to be used at all times, while the switches are achieved by a process of code rewiring (or code deformation), comprising a sequence of measurements and corresponding corrections.

Colladay and Mueller \cite{colladay_rewiring_2018} showed that during this switching process, a logical unitary can be implemented. While proving that the unitary must be Clifford, they did not show what Cliffords could be implemented, or provide a constructive method for creating a given Clifford.

In this paper, we focus on the special case where the initial and final codes are the same, and show how to generate the entire Clifford group within the code space. By appending to a code switching protocol, any Clifford operation can be implemented during the switch. Fault tolerance is immediate just by making the measurements themselves fault-tolerant by standard methods \cite{shor_fault-tolerant_1996}.

Our method has previously been described for the specific case of a single logical qubit in the surface code \cite{brown2017}, but we emphasise the broad applicability (which requires multi-qubit gates within a code block). For example, the 15-qubit Reed-Muller code combines this fault-tolerant implementation of the Clifford group with a transversal $T$ gate, enabling an alternative method to universal quantum computation from \cite{paetznick_universal_2013}. For any quantum Low Density Parity Check (LDPC) code \cite{breuckmann_quantum_2021}, where any error syndromes are measured on small blocks of qubits, the method preserves the LDPC property throughout.

\subsection{Code Rewiring} \label{sec:principles}

Suppose that $C$ is an $[[n,k,d]]$ stabilizer code, encoding $k$ logical qubits into $n$ physical qubits such that any logical operator has weight at least $d$. The stabilizers are Pauli strings of length $n$, drawn from the Pauli group $\mathcal{P}_n$. Key to the analysis of stabilizer codes are commutation relations between pairs of operators $A$ and $B$. We use a function $c(A,B)\in\{0,1\}$ to denote commutation properties:
$$AB=(-1)^{c(A,B)}BA.$$
For example, if our code has logical operators $L^{(i)}_X$ and $L^{(i)}_Z$ for $i=1,2,\ldots k$, they must satisfy commutation properties such as
$$c(L^{(i)}_X,L^{(j)}_Z)=\delta_{i,j},\qquad c(L^{(i)}_X,L^{(j)}_X)=0.$$

Code rewiring gives a sequence of measurements that can transform from one code into another. The principle behind code rewiring is established in Example 9 of \cite{gottesman_heisenberg_1998}, which we reformulate as the following statement.

\begin{lemma} \label{lem:basicprinciple}
Let $\ket{\psi}$ be a $+1$ eigenstate of the commuting operators $\{g_i\}_{i=1}^m\subset\mathcal{P}_n$. Let $g \in \mathcal{P}_{n}$ be an observable which anti-commutes with $g_{m}$. Suppose that we measure $g$ and apply the operator $g_{m}$ if the measurement outcome is $-1$. The operators are changed by:
\begin{itemize}
\item $g_{m}$ is replaced with $g$.
\item $g_i$ is replaced with $g_{m}^{c(g,g_i)}g_i$ for all $i\neq m$. 
\end{itemize}
\end{lemma}

This \emph{elementary code rewiring} was shown in \cite{colladay_rewiring_2018} to have the net effect of applying $U = (I + g g_{m})/\sqrt{2}$ to the state space. Moreover, $U$ is a Clifford operator, transforming Paulis to Paulis -- for any $\sigma\in\mathcal{P}_n$, $U\sigma U^\dagger=\sigma(gg_m)^{c(\sigma,gg_m)}\in\mathcal{P}_n$.

In \cite{colladay_rewiring_2018}, Colladay and Mueller use this procedure to provide an algorithm  to convert a state encoded in one stabilizer code $C$ into a state encoded in a different stabilizer code $C'$. Given sets of generators $g_{1}, \dots , g_{m}$ and $g'_{1}, \dots , g'_{m}$ for codes $C$ and $C'$, they find a sequence of elementary code rewirings which convert each $g_{j}$ into a corresponding $g'_{j}$, with the potential to induce a Clifford gate within the code space. The technique was demonstrated by mapping between the Steane code and the 15-qubit Quantum Reed-Muller code $QRM(4)$, while leaving open the question of which Clifford operators can be implemented, and how. %Earlier work \cite{anderson_fault-tolerant_2014} had also provided a map between these codes by considering both as different gauge-fixings of a larger code, similar to the approach taken in \cite{paetznick_universal_2013}. 

We will show that any Clifford operator can be implemented in this way. For this purpose, it is sufficient to consider the case in which $C=C'$: for any $V$ induced by a transformation from $C$ to $C'$, we simply implement $UV^\dagger$ on $C'$.

\section{A Rewiring Ancilla}

We are now going to give a construction that allows us to implement arbitrary Clifford gates on the logical space of the code $C$. To do so, we will identify one logical qubit which will initially be prepared in a Pauli basis state (the set $\{g_m\}$ of \cref{lem:basicprinciple} comprises both the code's stabilizers and the logical operator of that basis state). The output state of this ancilla is irrelevant, provided it is a known Pauli basis state. We can do this without loss of generality, although there are some codes for which this is more natural than others:
\begin{itemize}
\item An $[[n,k,d]]$ code with $k>1$ (such as the toric code, or the expected working regime of LDPC codes), always has an ancilla that can be used.
\item Subsystem codes \cite{poulin_stabilizer_2005}, such as the Reed-Muller code, have some gauge qubits inside the stabilizer. These are effectively logical operators of a code with more logical qubits fixed in a particular state. We can choose to use one of these.
\item A generic $[[n,1,\tilde d]]$ code $\tilde C$ can produce the $[[n,2,d]]$ code $C$ by removing any generator $g$, and using it instead as a logical operator, say $Z$. There is a corresponding logical $X$ as well\footnote{In general, the change in distance from $\tilde d$ to $d$ can be dramatic. Since $g$ becomes a new logical operator, $d\leq|g|$, which could be much smaller than $\tilde d$.%\todo{On the other hand, if we disregard errors on that logical qubit because it is a largely irrelevant gauge qubit, it is not as bad? Or just choose a new presentation of the group with one high weight stabiliser which we use}
}.
\end{itemize}
Having identified this ancilla, all measurements that we perform will commute with all the stabilizers of $C$. As such the code space is preserved, and the distance $d$ does not change. (Of course, in the $[[n,1,\tilde d]]$ case, removing a stabilizer from the code reduces the distance, $\tilde d>d$.)

We will now perform a protocol involving two elementary rewirings. Assume, without loss of generality, that the ancilla is the first logical qubit and is prepared in the $+1$ eigenstate of $L^{(1)}_Z$. First, we measure the system using
$$
g=L^{(1)}_YG
$$
where $G$ is a product of logical operators of the other qubits. We reiterate that this commutes with all the code stabilizers, so they don't change. By \cref{lem:basicprinciple}, we replace $L^{(1)}_Z\mapsto L^{(1)}_YG$, and then we have to work out how all the other logical operators change:
$$
L\mapsto {L^{(1)}_Z}^{c(G,L)}L.
$$
Now we perform a second measurement in the basis $L^{(1)}_X$. Again, the code space is unchanged. The ancilla qubit is returned in the $+1$ eigenstate of $L^{(1)}_X$, from which we could later start another round. The other logical states have transformed as
\begin{align}
L&\mapsto (L^{(1)}_YG)^{c(L^{(1)}_X,{L^{(1)}_Z}^{c(G,L)}L)}{L^{(1)}_Z}^{c(G,L)}L \nonumber\\
&=(L^{(1)}_YGL^{(1)}_Z)^{c(G,L)}L \nonumber\\
&=(iG)^{c(G,L)}L, \nonumber\\
&=\sqrt{G}L\sqrt{G}^\dagger \label{eq:unitary}
\end{align}
with the penultimate line following because the overall state is a $+1$ eigenstate of $L_X^{(1)}$, and the last line follows only up to an irrelevant global phase. The updates of these logical operators are sufficient to determine the unitary evolution of the subspace: $\sqrt{G}=e^{-i\pi/4}(\identity+iG)/\sqrt{2}$.

\subsection{Generating Cliffords}\label{sec:clifford}

Our goal is to show that any Clifford can be implemented by a sequence of code rewiring measurements. It is sufficient to demonstrate three unitaries: Hadamard, $S$ and controlled-\textsc{not} \cite{gottesman_theory_1998}. Equivalently, we can take $\sqrt{Y}$, $S$ and controlled-phase.

Single-qubit gates are now immediate from \cref{eq:unitary}: simply take $G=L^{(j)}_Z$ to create the $S$ gate, or $G=L^{(j)}_Y$ to create $\sqrt{Y}$.

For a two-qubit gate, let's take $G=L^{(2)}_ZL^{(3)}_Z$. We have that
$$
\sqrt{Z\otimes Z}=\text{diag}\left(\begin{bmatrix} 1 & -i & -i & 1 \end{bmatrix}\right).
$$
This is equivalent to the quantum circuit
$$
\begin{quantikz}
& \gate{S^\dagger} & \ctrl{1} & \\
& \gate{S^\dagger} & \ctrl{0} &
\end{quantikz}
$$
yielding the two-qubit gate required to completely generate the Clifford group within the code space\footnote{The Clifford group for \emph{all} logical qubits requires controlled-\textsc{not} between different code blocks, but all CSS codes have a transversal implementation of this. While it yields transversal c\textsc{not} at the logical level, these effects can be singled out by the single-qubit Cliffords that we have demonstrated here.}.

\subsection{Simplified Procedure}

While we have described the procedure for implementing the Clifford generators as a two-step process, they can be combined as a single circuit, with the corrections postponed to a single step. Furthermore, so long as we can keep track of the state of the ancilla, and so long as it remains in a Pauli basis state, it does not matter which of those states it finishes in. As such, any correction applied to the ancilla is unnecessary. Ultimately, this means that we can reduce the circuit to the one depicted in \cref{fig:circuit} (up to standard modifications that make the measurements fault tolerant \cite{shor_fault-tolerant_1996}).

\begin{figure}[!t]
\centering
\begin{adjustbox}{width=0.5\textwidth}
\tikzset{
operator/.style={draw,filling, inner sep=2pt,thickness,align=center,minimum width=0.7cm,minimum height=0.7cm}
}
\begin{quantikz}[wire types={q,q,q,b},classical gap vertical=0.07cm]
\lstick{\ket{0}} & \gate{H} &&\ctrl{2} & \gate{H} & \meter{}\wire[d][3]{c}\ar[r]{} & \rstick{$a$}\wireoverride{n} \\
\lstick{\ket{0}} & \gate{H} & \ctrl{2} && \gate{H} & \meter{}\ar[r]{} & \rstick{$b$}\wireoverride{n} \\
\lstick{$\ket{0}^{(1)}_L$} &&\gate{L_Y} & \gate{L_X} &&&\rstick{$L_Y^a\ket{+}_L^{(1)}$} \\
\lstick{$\ket{\psi}^{(2-k)}_L$} &&\gate{G}&&&\gate{G^{a\oplus b}} & \rstick{$U\ket{\psi}^{(2-k)}_L$}
\end{quantikz}
\end{adjustbox}
\caption{Circuit diagram for the implementation of a Clifford gate $U=\sqrt{G}$ on logical qubits 2 to $k$ of an error correcting code where all corrections have been combined into a single step.}\label{fig:circuit}
\end{figure}

When $a=b=0$, there is no change from our original protocol, implying that we have already proven that
$$
(\identity+L_X^{(1)})(\identity+L_Y^{(1)}G)\ket{0}^{(1)}\ket{\psi}=\ket{+}^{(1)}U\ket{\psi},
$$
up to normalisation. The correctness of \cref{fig:circuit} is then determined by verifying that
\begin{multline*}
G^{a\oplus b}(\identity+(-1)^aL_X^{(1)})(\identity+(-1)^bL_Y^{(1)}G)\ket{0}^{(1)}\ket{\psi} \\={L_Y^{(1)}}^a\ket{+}^{(1)}U\ket{\psi}.
\end{multline*}

\section{Summary}

We have shown that code rewiring can be used to generate any element of the Clifford group within the code space of a stabilizer code $C$, and that under some mild conditions, this can be carried out fault-tolerantly, with no cost to the distance of the code. This answers a question raised in \cite{colladay_rewiring_2018}. While our approach is efficient for certain non-trivial gates, it is unlikely that it will be the most efficient in general. As presented, combining two unitaries involves returning back to the original code as an intermediate step. For example, the construction of the two-qubit gate in \cref{sec:clifford} indicates that parallelising the procedure is not as simple as setting $G=L^{(2)}_ZL^{(3)}_Z$ to create $S^{(2)}S^{(3)}$. On the other hand, imagine that we want to implement both $\sqrt{G_1}$ and $\sqrt{G_2}$ where $c(G_1,G_2)=0$. If we have two ancilla qubits, then we can perform the two gates in parallel.
It would be interesting to determine whether there are more efficient sets of measurements.

Our approach provides a method to generate a set of gates which is universal for quantum calculations in the 15-qubit quantum Reed-Muller code $QRM(4)$, which is the smallest code to have a transversal $T$ gate \cite{koutsioumpas_smallest_2022}. We use just two rounds of measurements to implement a non-basis preserving Clifford operator, equivalent to Hadamard, where all other gates in the universal set can be implemented transversally. In contrast to the method in \cite{paetznick_universal_2013} in which the authors implement the Hadamard gate in $QRM(4)$ by applying Hadamard transversally to all physical qubits, making measurements and correcting, the code rewiring approach appears to require fewer gates and ancillas. Indeed, each elementary step is equivalent to the measurement of a stabilizer operator in Shor's method of error correction  \cite{shor_fault-tolerant_1996}, meaning that it could be incorporated directly into the error correction protocol. However, since Steane's method \cite{steane_efficient_1999}  is often used, a detailed comparison of the impact of that change would be required.

Quantum LDPC codes do not always present natural methods for performing logical operators (see for example \cite{breuckmann_quantum_2021}). The method we presented implements any Clifford operator while ensuring that the code remains an LDPC code throughout. Although the measurements may not be geometrically local, this is no more problematic than the stabilizers themselves not being geometrically local. Hence, this method provides a fault-tolerant implementation of Clifford operators on quantum LDPC codes. Supplemented by ancilla preparation, this can enable universal quantum computation. It should be noted, however, that the measurements must be high weight (at least $d$). This is necessarily the case for such a small number of steps. The main contribution of \cite{cohen_low-overhead_2022} is to provide a method for breaking down the measurement into multiple smaller steps via the introduction of ancilla qubits such that the operator to be measured is part of the stabilizer of the new, larger, LDPC code, and adding remarkably little to the required circuit depth. They also offer a promising analysis of the performance of such LDPC codes when implementing these gates.

\end{document}